\documentclass[twocolumn,twoside,10pt]{IEEEtran}
\usepackage{setspace}
\usepackage{cite,color}%
\usepackage{amsfonts}%
\usepackage{amsmath,url}%
\usepackage{amssymb}%
\usepackage{graphicx}
\usepackage{balance,epsfig}
\usepackage{amsmath,amssymb,graphicx,epsfig,url,cite,balance}
\begin{document}
\newcounter{eqncount}
\pagestyle{empty}
\thispagestyle{empty}

\title{Interference and Deployment Issues for Cognitive Radio Systems in Shadowing Environments}
\author{\authorblockA{Muhammad Fainan Hanif~\IEEEauthorrefmark{1}, Mansoor
Shafi~\IEEEauthorrefmark{2}, Peter J. Smith~\IEEEauthorrefmark{1}
and P.~Dmochowski~\IEEEauthorrefmark{3}}\\
\authorblockA{\IEEEauthorrefmark{1}Department of Electrical and Computer Engineering, University of Canterbury,
Christchurch, New Zealand}
\authorblockA{\IEEEauthorrefmark{2}~Telecom New Zealand, PO Box 293, Wellington, New
Zealand}\\
\authorblockA{\IEEEauthorrefmark{3}Faculty of Engineering, Victoria University of Wellington, New Zealand\\
Email:
mfh21@student.canterbury.ac.nz,~mansoor.shafi@telecom.co.nz,~p.smith@elec.canterbury.ac.nz,~pawel.dmochowski@vuw.ac.nz
}} \maketitle \thispagestyle{empty} \pagestyle{empty}
\begin{abstract}
In this paper we describe a model for calculating the aggregate
interference encountered by primary receivers in the presence of
randomly placed cognitive radios (CRs). We show that incorporating
the impact of distance attenuation and lognormal fading on each
constituent interferer in the aggregate, leads to a composite
interference that cannot be satisfactorily modeled by a lognormal.
Using the interference statistics we determine a number of key
parameters needed for the deployment of CRs. Examples of these are
the exclusion zone radius, needed to protect the primary receiver
under different types of fading environments and acceptable
interference levels, and the numbers of CRs that can be deployed. We
further show that if the CRs have \emph{apriori} knowledge of the
radio environment map (REM), then a much larger number of CRs can be
deployed especially in a high density environment. Given REM
information, we also look at the CR numbers achieved by two
different types of techniques to process the scheduling information.
\end{abstract}
\section{Introduction}
The conventional methodology adopted by spectrum regulatory agencies
is to grant exclusive licences to service providers to operate in a
particular frequency band. This inflexible approach has led to
severe under-utilization of the radio frequency (RF) spectrum.

Such observations have been strengthened by various measurement
campaigns \cite{report1}, \cite{report2} whose results have revealed
the surprising fact that the scarcity of spectrum is mainly due to
the fixed frequency allocation methodology.

The lack of efficient spectrum usage and consumers' ever-increasing
interest in wireless services have triggered a tremendous global
research effort on the concept of cognitive radios (CRs) or
secondary users (SUs). These CRs are deemed to be intelligent agents
capable of making opportunistic use of radio spectrum while
simultaneously existing with the legacy primary (licensed) users
(PUs) without harming their operation. The enormous interest in the
practical deployment of cognitive radios is reflected by the fact
that the IEEE has formed a special working group (IEEE 802.22) to
develop an air interface for opportunistic secondary access to the
TV spectrum.

In addition to ensuring quality of service (QoS) operation, the most
important and challenging task for the CRs is to avoid adverse
interference to the incumbent PUs. Accurate spectrum sensing
capabilities are being developed for CRs and it is envisaged that
either individually, or via collaboration, the secondary devices
will be able to detect the licensed users with a minimum probability
of failure. However, even with the best spectrum sensing techniques
the nature of the wireless channel will always result in some
interference at the PU due to the CRs. Hence, it is necessary to
characterize the nature of the interfering signals along with their
impact on the performance of the incumbent licensed users. In
addition, the amount of access that CRs are able to obtain without
too much impact on the PU is a key issue.

In this paper we focus on the issues described above. Via both
analysis and simulations we study the CR access problem and the
impact of CRs on the performance of the licensed devices. In
particular, we make the following contributions:
\begin{itemize}
\item Assuming lognormal shadowing and path loss effects we mathematically characterize
the cumulative distribution function (CDF) of the interference due
to the cognitive device. Further, we investigate the nature of the
distribution of the total interference \cite{Ghasemi} due to
multiple CRs.
\item In the same fading conditions we compute \emph{primary
exclusion zones} (PEZ) \cite{Mai1} in which CRs are not permitted to
operate. The PEZ approach allows access to CRs when the primary
device is willing to pay a price in the form of a reduction in its
threshold signal to noise ratio (SNR).
\item We determine the permissible number of CRs when
the \emph{radio environment map} (REM) \cite{J.Reed1} is
\emph{apriori} known to the CRs and we determine how these numbers
vary in the different fading environments.
\item Finally, we determine the scenarios (CR density, fading parameters) under
which REM based CR systems outperform PEZ based cognitive wireless
systems.
\end{itemize}
This paper is organized as follows: Section~II describes the system
model. Section~III derives the CDF of the interference seen by a PU
receiver and compares the analytical expression with simulations. We
also explore various parameters of the composite interference in a
CR network in this section. We introduce the PEZ and REM schemes in
Section~IV and compare their performance in Section~V. Finally, in
Section~VI we describe our conclusions.
\section{System Model}
\begin{figure}[t]
\centering
\includegraphics[width=0.75\columnwidth]{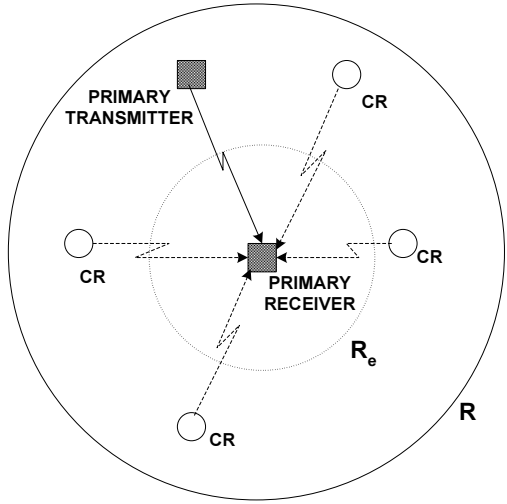}
\caption{System model ($R_0$ not shown)} \label{fig_1} \vspace{-3mm}
\end{figure}

Consider a PU receiver in the center of a circular region of radius
$R$. The PU transmitter is located uniformly in an annulus of outer
radius $R$ and inner radius $R_{0}$ centered on the PU receiver. It
is to be noted that we place the PU receiver at the center only for
the sake of mathematical convenience (see Fig. \ref{fig_1}). The use
of the annulus restricts devices from being too close to the
receiver. This matches physical reality and also avoids problems
with the classical inverse power law relationship between signal
strength and distance \cite{Mai}. In particular, having a minimum
distance, $R_0$, prevents the signal strength from becoming infinite
as the transmitter approaches the receiver. Similarly, we assume
that multiple CR transmitters are uniformly located in the annulus.
At any given time, each CR has a probability of seeking a
connection, given by the activity factor, $p$. The number of CRs
wishing to operate is denoted $N_{CR}$. Of these CRs, a certain
number will be accepted depending on the allocation mechanism.
Hence, a random number of CRs denoted $N\leq N_{CR}$ will transmit
during the PU transmission and create interference at the PU
receiver.

The received signal strength for both the PU transmitter to PU
receiver and CR transmitter to PU receiver is assumed to follow the
classical distance dependent, lognormal shadowing model. For a
generic interferer, this is given by
\begin{equation}\label{first}
I =BLr^{-\gamma}=B10^{\tilde{X}/10}r^{-\gamma}= Be^Xr^{-\gamma}
\end{equation}
where $r$ is the random distance from the transmitter to the
receiver, $\gamma$ is the path loss exponent (normally in the range
of 2 to 4) and $L$ is a shadow fading variable. The lognormal
variable, $L$, is given in terms of the zero mean Gaussian,
$\tilde{X}$, which has standard deviation $\sigma$ (dB) and
$X=\beta\tilde{X}$ where $\beta=\ln(10)/10$. The standard deviation
of $X$ is denoted by $\sigma_{x}$. The constant $B$ is determined by
the transmit power. The desired primary signal strength, $S$, has
the same form, with a different transmit power, so that
$S=AL_pr_p^{-\gamma}$. Note that all the links are assumed to be
independent and identically distributed (i.i.d.) so that spatial
correlation is ignored. The results in the paper can be further
generalized to more realistic scenarios by considering the system
models such as those presented in \cite{Revsug1,Revsug2}.
\section{Statistical Characterization of Interference at the Primary Receiver}
In this section we investigate the interference at the PU receiver
due to one or more CRs. Firstly we characterize the interfering
signal, given in (\ref{first}), by computing the CDF,
$F_{I}(x)=P(I<x)$. This can be done as follows:
\setlength{\arraycolsep}{0.0em}
\begin{eqnarray}\label{ninth}
F_{I}(x)&{}={}&P\big(Be^{X}r^{-\gamma}<x\big)\nonumber\\
&{}={}&P\Bigg(e^{-X}r^{\gamma}>\frac{B}{x}\Bigg)\nonumber\\
&{}={}&P(e^{U}r^{\gamma}>y)\nonumber\\
&{}={}&E_{U}\Big(P\Big(r>y^{\frac{1}{\gamma}}e^{\frac{-U}{\gamma}}\Big)\Big)\nonumber\\
&{}={}&E_{U}\Big(1-F_R\Big(y^{\frac{1}{\gamma}}e^{\frac{-U}{\gamma}}\Big)\Big)
\end{eqnarray}
\setlength{\arraycolsep}{5pt}
\hspace{-1.5mm}In (\ref{ninth}), $U=-X$, $E_U$ represents
expectation over the random variable $U$ and $y=B/x$. To evaluate
the expectation in (\ref{ninth}) we note that the CDF of $r$ is
given by:
\begin{displaymath}
F_{R}(r)=\frac{r^2-R_0^2}{R^2-R_0^2},\qquad R_0\leq r\leq R.
\end{displaymath}
Using this CDF, (\ref{ninth}) can be rewritten as:
\begin{equation}\label{newnew6}
F_{I}(x)=E_U(G_R(y)),
\end{equation}
where
\begin{equation}\label{newnew7}
G_{R}(y) = \left\{ \begin{array}{ll}
0 & \textrm{$u<w_0$}\\
\bigg(\frac{R^2-y^{\frac{2}{\gamma}}e^{\frac{-2u}{\gamma}}}{R^2-R_0^2}\bigg) & \textrm{$w_0<u<w_1$}\\
1 & \textrm{$u>w_1$}
\end{array} \right.
\end{equation}
and $w_{0}=\ln(yR^{-\gamma})$, $w_{1}=\ln(yR_{0}^{-\gamma})$. Since
$U=-X$ is Gaussian, $U\sim\mathcal{N}(0,\sigma_{x}^2)$, the CDF,
$F_I(x)=E_U(G_R(y))$, becomes:
\setlength{\arraycolsep}{0.0em}
\begin{eqnarray}\label{ten}
F_{I}(x)&{}={}&\int_{w_0}^{w_1}\!\!\!\frac{R^2-y^{\frac{2}{\gamma}}e^{\frac{-2u}{\gamma}}}{R^2-R_0^2}f(u)du+\int_{w_1}^{\infty}\!\!f(u)du\nonumber\\
&{}={}&\frac{R^2}{(R^2-R_0^2)\sqrt{2\pi
\sigma_{x}^2}}\int_{w_0}^{w_1}\!\!\!e^{\frac{-u^2}{2\sigma_{x}^2}}du\nonumber\\
&&{-}\:\frac{(\frac{B}{x})^{2/\gamma}}{(R^2-R_0^2)\sqrt{2\pi
\sigma_{x}^2}}\int_{w_0}^{w_1}\!\!\!e^{\frac{-2u}{\gamma}-\frac{-u^2}{2\sigma_{x}^2}}du\nonumber\\
&&{+}\:\frac{1}{\sqrt{2\pi
\sigma_{x}^2}}\int_{w_1}^{\infty}\!\!\!e^{\frac{-u^2}{2\sigma_{x}^2}}du.
\end{eqnarray}
\setlength{\arraycolsep}{5pt}

\hspace{-2.9mm}All the integrals in (\ref{ten}) can be written in
terms of integrals of Gaussian PDFs. Hence (\ref{ten}) can be
written as shown in (\ref{eleven}), where $F_Z(.)$ is the CDF of a
standard Gaussian.
\begin{figure*}[t]
\normalsize \setcounter{eqncount}{\value{equation}}
\setcounter{equation}{5}
\begin{align}
\label{eleven} F_{I}(x)&
=1-F_Z\Bigg(\frac{w_1}{\sigma_{x}}\Bigg)+\frac{1}{R^2-R_0^2}\Bigg[\Bigg\{R^2
F_Z\Big(\frac{w_1}{\sigma_{x}}\Big)-R^2F_Z\Big(\frac{w_0}{\sigma_{x}}\Big)\Bigg\}\\\nonumber&
-\Big(\frac{B}{x}\Big)^{\frac{2}{\gamma}}e^{\big(\frac{2\sigma_{x}^2}{\gamma^2}\big)}
\Bigg\{F_Z\Bigg(\frac{w_1+2\sigma_{x}^2/\gamma}{\sigma_{x}}\Bigg)-F_Z\Bigg(\frac{w_0+2\sigma_{x}^2/\gamma}{\sigma_{x}}\Bigg)\Bigg\}\Bigg]
\end{align}
\setcounter{equation}{\value{eqncount}} \hrulefill
\end{figure*}
\addtocounter{equation}{1}

\begin{figure}[t]
\centering
\includegraphics[width=0.95\columnwidth]{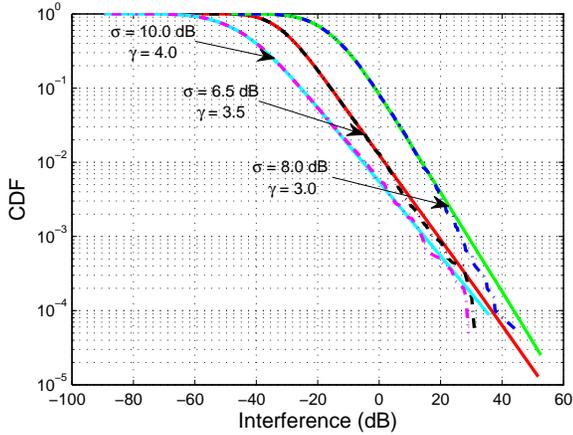}
\caption{A comparison of analytical and simulated CDFs of
interference over a range of propagation parameters.} \label{fig_2}
\end{figure}

In order to validate the CDF given in (\ref{eleven}), we compare the
results of this expression with the CDF obtained using Monte Carlo
simulation as shown in Fig. \ref{fig_2}. It can be seen that the
analytical and simulated interference CDFs agree over a wide range
of propagation parameters. The discrepancy between the analytical
and simulated curves in the tail region is due to the limited number
of realizations used and the inherent long tailed properties of the
lognormal. Hence, a complete characterization of the interference
due to a single CR is possible. Next, we consider multiple CRs.
In cognitive wireless networks the PU device under consideration may
be affected by the interference due to many CRs. In this case, the
total interference, denoted $I_{TOT}$, is given by:
\begin{equation}\label{second}
I_{TOT}=\sum_{i=1}^{N}Be^{X_{i}}r_{i}^{-\gamma}=\sum_{i=1}^{N}BL_{i}r_{i}^{-\gamma},
\end{equation}
where the parameters are as defined in (\ref{first}). Equation
(\ref{second}) is a random sum of a finite number of lognormals with
each lognormal being multiplied by a random distance factor.
Problems similar to this, but involving a non random sum without
incorporating the random distance factor, have been tackled in the
past \cite{B-Abu1,B-Abu,B-Xie,molisch,Renzo,Liu1}. Historically,
lognormal approximations to this summation have been envisaged.
Approximations are definitely required since although (\ref{eleven})
gives an analytical result for a single interferer, it is too
complex to allow an exact approach for sums of interferers. Hence
one is tempted to use the same lognormal approximation for
(\ref{second}). However, we show that lognormal approximations are
not accurate. For convenience, consider a lognormal approximation to
a single interferer, $I$, of the form given in (\ref{first}). Let
the moments of $I$ be denoted by $E(I^j)=m_j$. We seek to
approximate $I$ with the lognormal $Y=e^Z$ where $Z\sim
\mathcal{N}(\mu_z,\sigma_z^2)$. The simplest approach to fitting $Y$
is via the Fenton-Wilkinson approximation \cite{F-W,B-Abu1} which
computes the first two moments, so that:
\begin{equation}\label{p3}
E(Y^k)=E(I^k),\quad k=1,2
\end{equation}
Hence, the lognormal approximation has perfect moments up to order
2. To demonstrate the lack of fit we consider the skewness of $Y$
and $I$ which also involves the third moment. For any lognormal, say
$Y$, the third moment is related to the first two by:
\begin{equation}\label{p4}
E(Y^3)=\Bigg(\frac{E(Y^2)}{E(Y)}\Bigg)^3.
\end{equation}
Now, consider the skewness of $I$,
\setlength{\arraycolsep}{0.0em}
\begin{eqnarray}\label{newnew3}
SK(I)&{}={}&\frac{E[(I-m_{1})^3]}{(m_2-m_1^2)^{3/2}}\nonumber\\
&{}={}&\frac{m_3+2m_1^3-3m_1m_2}{(m_2-m_1^2)^{3/2}}.
\end{eqnarray}
\setlength{\arraycolsep}{5pt}
\hspace{-1.5mm}Similarly, the skewness of $Y$ can be written as:
\setlength{\arraycolsep}{0.0em}
\begin{eqnarray}\label{p5}
SK(Y)&{}={}&\frac{E(Y^3)+2E(Y)^3-3E(Y)E(Y^2)}{(E(Y^2)-E(Y)^2)^{3/2}}\nonumber\\
&{}={}&\frac{E(Y^3)+2m_1^3-3m_1m_2}{(m_2-m_1^2)^{3/2}}
\end{eqnarray}
\setlength{\arraycolsep}{5pt}
Hence, the lognormal approximation is more skewed than the real
interference if $E(Y^3)>m_3$. Now, consider
%
%
\vspace{-3mm} \setlength{\arraycolsep}{0.0em}
\begin{eqnarray}\label{p6}
\frac{E(Y^3)}{m_3}&{}={}&\frac{(m_2/m_1)^3}{m_3}\nonumber\\
&{}={}&\frac{\bigg(\frac{B^2E(e^{2X})E(r^{-2\gamma})}{BE(e^{X})E(r^{-\gamma})}\bigg)^3}{B^3E(e^{3X})E(r^{-3\gamma})}\nonumber\\
&{}={}&\frac{\bigg(\frac{E(e^{2X})}{E(e^{X})}\bigg)^3}{E(e^{3X})}\frac{\bigg(\frac{E(r^{-2\gamma})}{E(r^{-\gamma})}\bigg)^3}{E(r^{-3\gamma})}\nonumber\\
&{}={}&\frac{\bigg(\frac{E(r^{-2\gamma})}{E(r^{-\gamma})}\bigg)^3}{E(r^{-3\gamma})}.
\end{eqnarray}
\setlength{\arraycolsep}{5pt}
\hspace{-1.5mm}The above results
follow since (\ref{p4}) also holds for the lognormal $e^{X}$. The
moments of $r$ in (\ref{p6}) can be found as:
\begin{equation}\label{p7}
E(r^{-k\gamma})=\frac{2(R^{2-k\gamma}-R_{0}^{2-k\gamma})}{(R^2-R_0^2)(2-k\gamma)}.
\end{equation}
Equation (\ref{p7}) is obtained using $f(r) =
\frac{2r}{(R^2-R_0^2)}$ as the PDF for the CR distance in the
annulus $R_0 < r < R$. For large $R$ (1000 m in our case), small
$R_0$ (we use 1m) and with $\gamma\geq3$, (\ref{p7}) gives:
\begin{equation}\label{p8}
E(r^{-k\gamma})\approx\frac{-1}{R^2}\bigg(\frac{2}{2-k\gamma}\bigg)=\frac{2}{R^2(k\gamma-2)}.
\end{equation}
Thus, the ratio of the moments in (\ref{p6}) becomes:
\setlength{\arraycolsep}{0.0em}
\begin{eqnarray}\label{p9}
\frac{E(Y^3)}{m_3}&{}\approx{}&\bigg(\frac{\gamma-2}{2\gamma-2}\bigg)^3\frac{R^2(3\gamma-2)}{2}
\end{eqnarray}
\setlength{\arraycolsep}{5pt}
\hspace{-1.5mm}It can be observed from the above expression that for
typical values of the parameters, $E(Y^3)>>m_3$, and the ratio is of
the order of $R^2$. Thus the equivalent lognormal will be massively
more skewed than the real interference. Now as skewness is a key
shape determining factor (especially in the tails), the simple
lognormal approximation will not be accurate. Note that this large
discrepancy in skewness is due to the random distance factors.
Exactly the same conclusions are reached when attempting to fit sums
of interferers. Hence, it appears that a simple lognormal
approximation will not suffice and further research is required. In
one possible approach \cite{Ghasemi} has approximated the total
interference of the form given in (\ref{second}) with a more
flexible shifted \emph{three parameter} lognormal random variable
using cumulant matching. Comparison between the simulation results
and this approximation shows good matching especially in the head
portion of the CDF. Similarly, the recent work in \cite{Revsug3}
analyzes the aggregate interference in the absence of any shadow
fading. This makes it difficult to compare the two sets of results.
In this paper, we use simulations to assess the REM and PEZ schemes
as discussed in Section~IV.
\section{REM And PEZ  Based Schemes}
In all simulations CRs are located uniformly in the primary coverage
area. The number of active CRs, $N_{CR}$, is binomially distributed
with a maximum number of CRs given by $\pi R^2D_{CR}$, where
$D_{CR}$ is the density of the CRs (number of CRs per $\text{m}^2$)
and we ignore the negligible hole in the circle of radius $R_0$. The
binomial probability that a CR wishes to transmit is given by the
activity factor, $p=0.1$. The primary receiver is at the center of
the coverage area and the primary transmitter is also uniformly
located in the primary coverage area. In this section we consider
two fundamentally different schemes for managing the interference at
the PU receiver. The REM approach utilizes instantaneous knowledge
of all the interference values whereas the PEZ approach only uses
average information. Assume there are $N_{CR}$ CR transmitters which
desire a connection. Each of the $N_{CR}$ CRs has an interference
power at the primary receiver given by (\ref{first}) and denoted
$I_1, I_2,\ldots,I_{N_{CR}}$. Based on these interference values,
the REM and PEZ approaches are described below.
\subsection{REM Approach}
The REM approach \cite{J.Reed1} assumes that $I_1,
I_2,\ldots,I_{N_{CR}}$ are known and selects those CRs for
transmission which satisfy an SINR constraint. The constraint chosen
is that the added interference must not decrease the SNR by more
than $\Delta$ dB. For example, if SNR = 10 dB in the absence of CRs,
then those CRs chosen must give $\text{SINR}\geq (10-\Delta)$ dB.
Two methods are chosen for selection, a centralized approach and a
decentralized approach.
\subsubsection{Centralized Selection}
Here we assume that a centralized controller knows $I_1,
I_2,\ldots,I_{N_{CR}}$ instantaneously and creates a list of the
ordered interferers as $I_{(1)}\leq I_{(2)}\leq \ldots \leq
I_{(N_{CR})}$. The first $n$ CRs are selected such that
$\sum_{i=1}^{n}I_{(i)} \leq \Delta$ dB and $\sum_{i=1}^{n+1}I_{(i)}
> \Delta$ dB.

\subsubsection{Decentralized Selection}
Here we assume that the CRs are considered in their original order
which can be interpreted as their order of arrival. Each interferer
is considered in turn and is accepted if the combined interference
from previously accepted CRs and the current CR is less than
$\Delta$ dB. If a CR is not accepted, the next CR in the list is
investigated.
\subsection{PEZ Approach}
The REM approach uses a detailed REM to give information about the
interference resulting from any CR. In contrast, the PEZ approach
\cite{Mai1} only uses location information to control the access of
CRs. A simple exclusion zone is created with radius $R_e$ around the
primary receiver. No CR is allowed to transmit inside the PEZ and
all CRs outside the PEZ are permitted. The radius, $R_e$, is set so
that the SINR within the PU coverage area is degraded by a certain
amount. Specifically, the primary coverage area is defined to give
an SNR greater than 5 dB, 95\% of the time. By allowing CRs to
operate we accept a new SINR target, less than 5 dB, which is
achieved at least 95\% of the time.
\section{Simulation Results}
\begin{figure}[t]
\centering
\includegraphics[width=0.97\columnwidth]{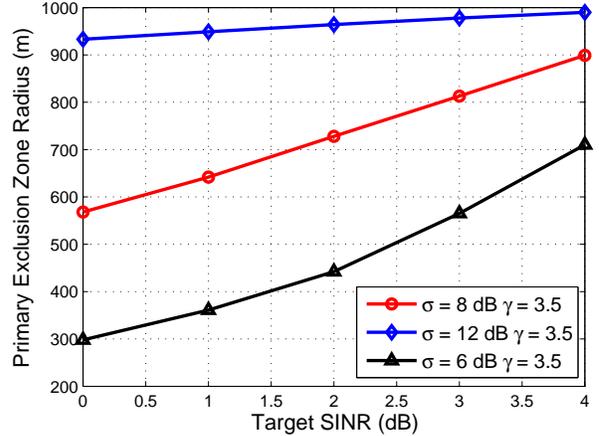}
\caption{The effect of $\sigma$ and the target SINR on the PEZ
radius for a medium density of CRs.} \label{fig1}
\end{figure}
\begin{figure}[t]
\centering
\includegraphics[width=0.95\columnwidth]{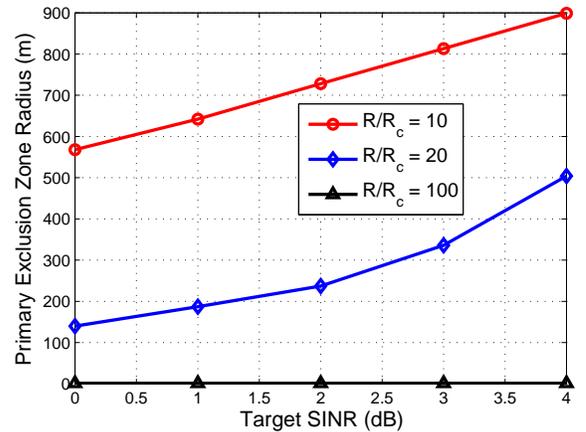}
\caption{PEZ radius vs target SINR for different values of the ratio
of primary to secondary device coverage areas ($\sigma=8$ dB,
$\gamma=3.5$).} \label{fig2}
\end{figure}
\begin{figure}[t]
\centering
\includegraphics[width=0.95\columnwidth]{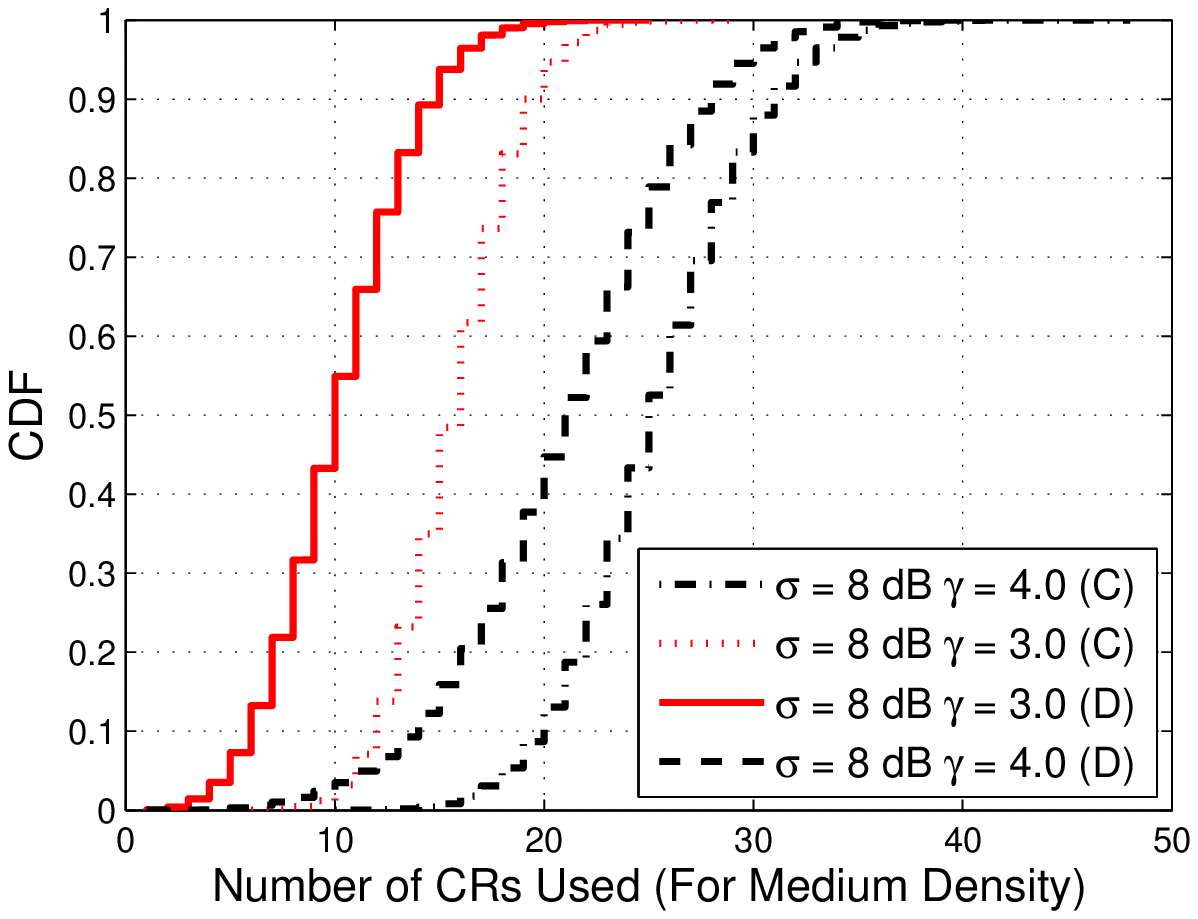}
\caption{CDF of the number of CRs obtained using REM based
approaches for various $\gamma$ values. D and C denote decentralized
and centralized approaches.} \label{fig3}
\end{figure}
\begin{figure}[t]
\vspace{0.0mm} \centering
\includegraphics[width=0.95\columnwidth]{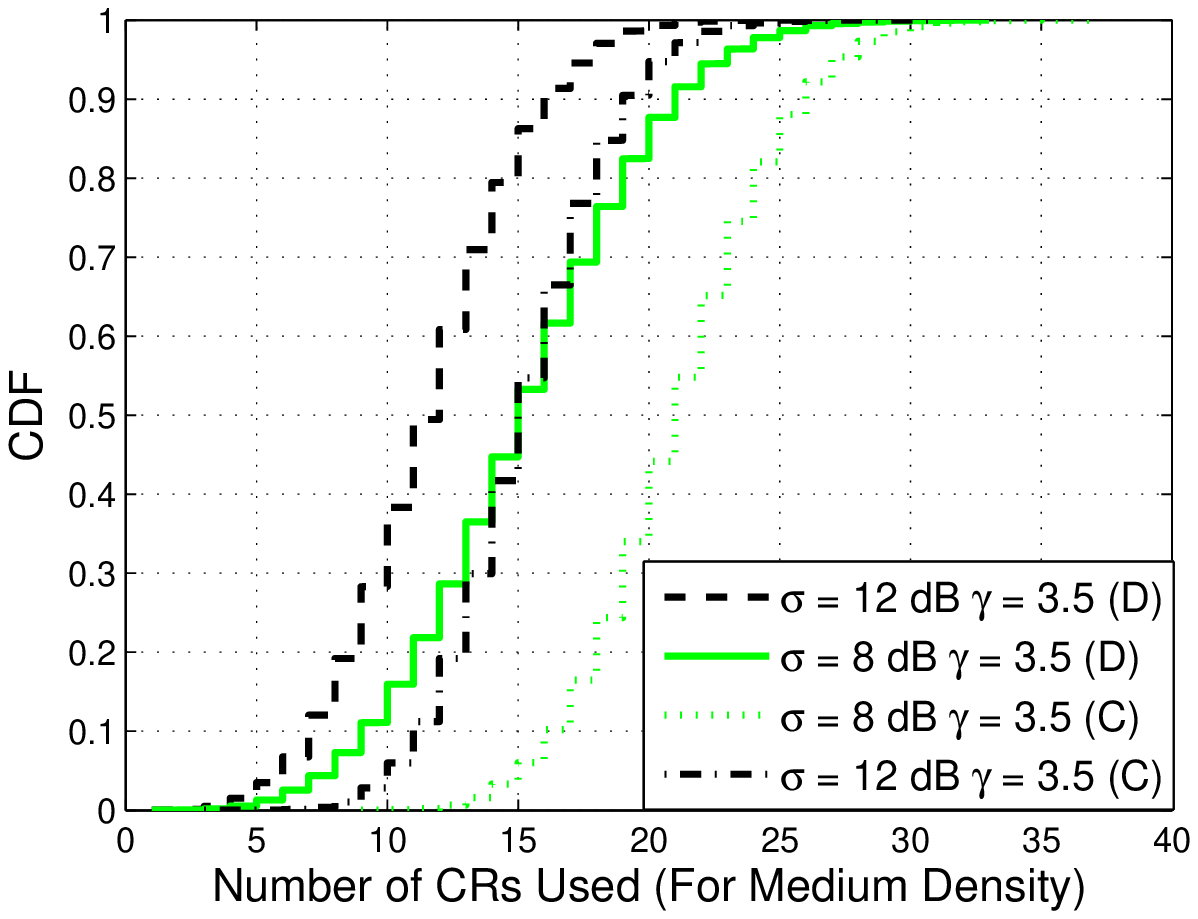}
\caption{CDF of the number of CRs obtained using REM based
approaches for various $\sigma$ values. D and C denote decentralized
and centralized approaches.} \label{fig4}
\end{figure}
\begin{figure}[t]
\centering
\includegraphics[width=0.93\columnwidth]{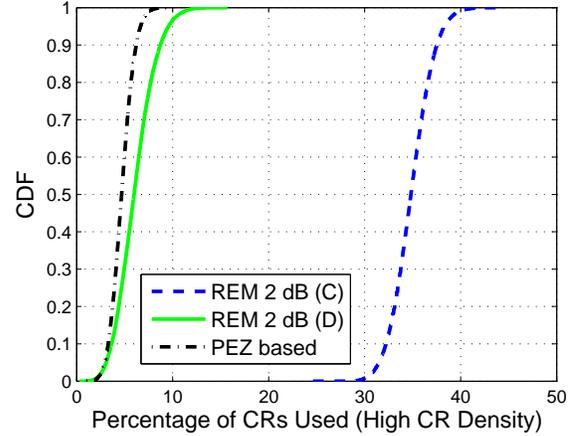}
\caption{Percentage of CRs given access for a high CR density
($\sigma=8$ dB, $\gamma=3.5$).} \label{fig5}
\end{figure}
\begin{figure}[t]
\vspace{-0 mm}\centering
\includegraphics[width=0.95\columnwidth]{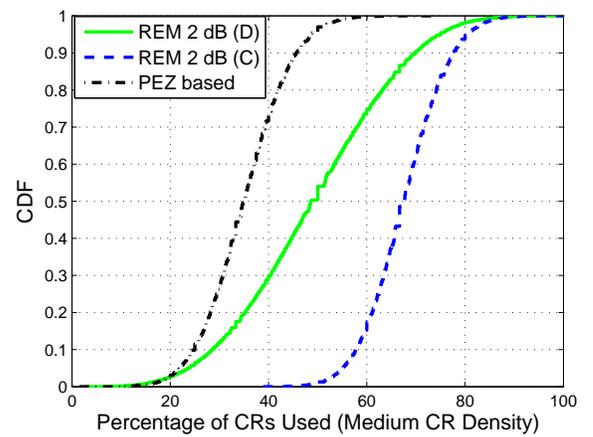}
\caption{Percentage of CRs given access for a medium CR density
($\sigma=8$ dB, $\gamma=3.5$).} \label{fig6}
\end{figure}
We assume a PU coverage radius of $R=1000$ m, and the transmit power
is adjusted such that the SNR in the coverage area exceeds 5 dB with
probability 0.95. In other words the area reliability with a 5 dB
target is 95\%. The CR transmit power is also chosen so that 5 dB is
achieved at least 95\% of the time for a given CR coverage radius,
$R_c$. We take $R_0=1$ m. Two kinds of CR penetration densities were
chosen, a high density of 10,000 CRs per sq. km and a corresponding
moderate density of 1000 CRs per sq. km. Additionally, it was
assumed that only 10\% of the CRs wish to be active at any one time.
The values of the propagation constants, $\gamma$ and $\sigma$ are
given on the relevant figures.
\subsection {Exclusion Zone Results}
Given a variety of target SINRs, Fig. \ref{fig1} shows the PEZ
radius for different values of $\sigma$. For example, if the
interference degrades the target SNR from 5 dB to an SINR of 4 dB,
then the PEZ radius is approximately 700 m, for $\sigma=6$ dB and
$\gamma=3.5$. It is interesting to note that the PEZ radius excludes
virtually the entire PU coverage area for all target SINRs in [0-5]
dB when $\sigma = 12$ dB, corresponding to dense urban areas. This
result implies that for a given target SINR, environments with
larger $\sigma$ will result in higher interference and an increased
$R_e$. This observation is consistent with previous observations
reported in \cite{Viterbi}.

Increasing the CR transmit power or increasing $R_c$ will
correspondingly increase the interference and hence the PEZ radius.
Figure~\ref{fig2} shows the PEZ radius vs target SINR for three
different values of $R/R_c$. These calculations were done for
$\sigma = 8$ dB and $\gamma = 3.5$. Reducing the CR transmit power
will obviously result in a lower PEZ radius.
\subsection {Numbers of CRs}
Figures \ref{fig3} and \ref{fig4} show CDFs of the number of CRs for
the two different types of REM approach and the impact of varying
the fading parameters. In both these figures, the centralized
approach is superior, since it is designed to pick up the maximum
number of CRs that aggregate to make up the acceptable interference
degradation. We also note from Fig. \ref{fig3} that increasing
$\gamma$ increases the number of permissible CRs.  This is because
environments where $\gamma =4$ will experience less interference
compared to environments where $\gamma =3$ due to increased path
loss. Looking at Fig.~\ref{fig4}, increasing $\sigma$ decreases the
permissible number of CRs. This result reinforces the conclusion of
Fig.~\ref{fig1} where increasing $\sigma$ increased the PEZ radius -
effectively reducing the area in which CRs operate and also reducing
the permissible number of CRs. Note that dense urban environments
are characterized by $\gamma$ values of 4 and above and $\sigma$
values of 8 dB and above. These two parameters have opposing effects
on the permissible number of CRs.

Figure~\ref{fig5} compares the PEZ and REM approaches in terms of
the percentage of CRs that gain access in a high density
environment. The centralized approach is far superior, showing the
advantage gained if the CR knows the radio environment. This
advantage is dissipated by the decentralized approach as effectively
a few CRs consume the permissible interference budget (2 dB in this
case). The PEZ approach is worse than the decentralized strategy. It
is an important result that the decentralized REM approach, which
can be thought of as a first-come-first-served mechanism, results in
better access for the CRs than the PEZ approach. Hence, the overhead
of obtaining the REM can result in improved access. It is critical
that the REM information be used in an intelligent allocation
process. Figure~\ref{fig6} revisits the results in Fig.~\ref{fig5}
for a lower CR density. Here too the centralized approach is better,
but now the decentralized approach shows an even bigger advantage
over the PEZ approach for higher values of the CDF.

The results in Figs. 5-8 taken collectively are the key contribution
of this paper. They clearly show the advantage in terms of
permissible CR numbers if a knowledge of the radio environment is
made available to the CRs. Furthermore, they show that the full REM
gains are only obtained if a smart access control algorithm is used
which chooses many CRs with low interference instead of a few
stronger interferers which might subsume the interference budget.
\section{Conclusion}
The interference due to a single CR can be characterized in closed
form for the scenario considered. However, the total interference
due to multiple CRs is more difficult. Simple lognormal
approximations are shown to be inaccurate and more complex models
are required. Two interference management approaches have been
considered based on REM and PEZ ideas. The REM approach requires
considerable higher overheads but can perform substantially better
than the PEZ approach. To achieve substantial gains an intelligent
allocation method is, however, essential. \balance

\bibliographystyle{IEEEtran}
\bibliography{IEEEabrv,interference}
\end{document}